\documentclass[twocolumn]{aastex63} %linenumbers
\usepackage{amsmath,amstext}
\usepackage{balance}
\usepackage{color}
\usepackage{comment}

\graphicspath{{./}{figures/}}
\begin{document}

\title{Accreting on the edge: a luminosity-dependent cyclotron line in the Be/X-ray Binary 2S 1553-542 accompanied by accretion regimes transition}

\correspondingauthor{cmalacaria@usra.edu}

%\tiny{  & ($d,\dot{M}$ fix, $B$ free) & ($d,B,\dot{M}$ fix) & ($d,B$ fix, $\dot{M}$ free)} \\

\author[0000-0002-0380-0041]{C.~Malacaria}
\affiliation{Universities Space Research Association, Science and Technology Institute, 320 Sparkman Drive, Huntsville, AL 35805, USA}
\affiliation{NASA Marshall Space Flight Center, NSSTC, 320 Sparkman Drive, Huntsville, AL 35805, USA}\thanks{NASA Postdoctoral Fellow}

\author{Y.~Bhargava}
\affiliation{Inter-University Centre for Astronomy and Astrophysics, Post box no. 4, Ganeshkhind, Pune-411007 India}

\author[0000-0001-7532-8359]{Joel~B.~Coley}
\affiliation{Department of Physics and Astronomy, Howard University, Washington, DC 20059, USA}
\affiliation{CRESST and Astroparticle Physics Laboratory, NASA Goddard Space Flight Center, Greenbelt, MD 20771, USA}

\author{L.~Ducci}
\affiliation{Institut f\"ur Astronomie und Astrophysik, Kepler Center for Astro and Particle Physics, Universit\"at T\"ubingen, Sand 1, 72076 T\"ubingen, Germany}

\author[0000-0002-1131-3059]{P.~Pradhan}
\affiliation{Massachusetts Institute of Technology, 77 Massachusetts Ave., Cambridge, MA 02139, USA}

%\author{D.~Bhattacharya}
%\affiliation{?}

\author{R.~Ballhausen}
\affiliation{Department of Astronomy, University of Maryland, College Park, MD 20742, USA}
\affiliation{NASA-GSFC/CRESST, Astrophysics Science Division, Greenbelt, MD 20771, USA}

\author{F.~Fuerst}
\affiliation{Quasar Science Resources S.L for European Space Agency (ESA), European Space Astronomy Centre (ESAC), Camino Bajo del Castillo s/n, 28692 Villanueva de la Cañada, Madrid, Spain}

\author{N.~Islam}
\affiliation{Center for Space Science and Technology, University of Maryland, Baltimore County, 1000 Hilltop Circle, Baltimore, MD 21250, USA}
\affiliation{X-ray Astrophysics Laboratory, NASA Goddard Space Flight Center, Greenbelt, MD 20771, USA}

\author[0000-0002-6789-2723]{G.~K.~Jaisawal}
\affil{National Space Institute, Technical University of Denmark, Elektrovej 327-328, DK-2800 Lyngby, Denmark}

\author{P. Jenke}
\affiliation{University of Alabama in Huntsville (UAH), Center for Space Plasma and Aeronomic Research (CSPAR), 301 Sparkman Drive, Huntsville, Alabama 35899}

\author[0000-0001-9840-2048]{P.~Kretschmar}
\affiliation{European Space Agency (ESA), European Space Astronomy Centre (ESAC), Camino Bajo del Castillo s/n, 28692 Villanueva de la Cañada, Madrid, Spain}

\author{I.~Kreykenbohm}
\affiliation{Remeis-Observatory and Erlangen Centre for Astroparticle Physics, Friedrich-Alexander-Universit\"at Erlangen-N\"urnberg, Sternwartstr.~7, 96049 Bamberg, Germany}

\author[0000-0002-4656-6881]{K.~Pottschmidt}
\affiliation{CRESST and Astroparticle Physics Laboratory, NASA Goddard Space Flight Center, Greenbelt, MD 20771, USA}
\affiliation{Department of Physics and Center for Space Science and Technology, University of Maryland, Baltimore County, Baltimore, MD
21250, USA}

\author{E.~Sokolova-Lapa}
\affiliation{Remeis-Observatory and Erlangen Centre for Astroparticle Physics, Friedrich-Alexander-Universit\"at Erlangen-N\"urnberg, Sternwartstr.~7, 96049 Bamberg, Germany}

\affiliation{Sternberg Astronomical Institute, M.~V.~Lomonosov Moscow State University, Universitetskij pr., 13, Moscow 119992, Russia}

\author{R.~Staubert}
\affiliation{Institut f\"ur Astronomie und Astrophysik, Kepler Center for Astro and Particle Physics, Universit\"at T\"ubingen, Sand 1, 72076 T\"ubingen, Germany}

\author[0000-0003-2065-5410]{J.~Wilms}
\affiliation{Remeis-Observatory and Erlangen Centre for Astroparticle Physics, Friedrich-Alexander-Universit\"at Erlangen-N\"urnberg, Sternwartstr.~7, 96049 Bamberg, Germany}

\author[0000-0002-8585-0084]{C.A.~Wilson-Hodge}
\affiliation{ST 12 Astrophysics Branch, NASA Marshall Space Flight Center, Huntsville, AL 35812, USA}

\author[0000-0002-4013-5650]{Michael T.~Wolff}
\affil{Space Science Division, U.S. Naval Research Laboratory, Washington, DC 20375-5352, USA}

\begin{abstract}
Accreting X-ray pulsars (XRPs) undergo luminous X-ray outbursts during which the luminosity-dependent spectral and timing features of the neutron star's emission can be analyzed in detail, thus shedding light on the accretion regime at work.
We took advantage of a monitoring campaign performed with \textit{NuSTAR}, \textit{Swift}/XRT, \textit{AstroSat} and \textit{NICER}, to follow the Be/X-ray Binary 2S 1553-542 along one of its rare outbursts and trace its spectral and timing evolution.
We report the discovery of a luminosity-dependent cyclotron line energy for the first time in this source. 
The pulse profiles and pulsed fraction
%Other spectral parameters, as well as the timing properties of the source, 
also show variability along the outburst, consistently with the interpretation that the source transitions from the sub-critical to the super-critical accretion regime, separated by a critical luminosity of $L_{\rm crit}\approx4\times10^{37}\,$erg/s.
\end{abstract}


\section{Introduction}\label{sec:introduction}

Accreting X-ray pulsars (XRPs) are binary systems where a neutron star (NS) accretes matter supplied by a donor companion star via stellar wind or Roche-lobe overflow.
Most of them pertain to the subclass of Be/X-ray Binaries (BeXRBs), in which the donor companion is a B star whose circumstellar decretion disk shows H$\alpha$ Balmer emission lines.

2S 1553-542 is a BeXRB discovered with the \textsl{SAS-3} observatory \citep{Walter76}, which also detected pulsations at about 9.3 s several years later \citep{Kelley82}.
Only a few X-ray outbursts have been observed from this source (see \citealt{Tsygankov16}, and references therein).
The optical companion has been identified as a B1-2V type star \citep[]{Lutovinov16}, and the system lies at a distance of $20\pm4\,$kpc \citep[]{Tsygankov16}.
No Gaia counterpart is found in the Early Data Release 3 \citep{Fabricius21} within
%\textbf{
${\sim}5.5^{\arcsec}$ from the SIMBAD astronomical database \citep[]{Wenger00} source position,  nor from the \textit{Chandra} source position determined by \citet[]{Lutovinov16}.
%}

\begin{figure}[!t]
\includegraphics[width=.47\textwidth]{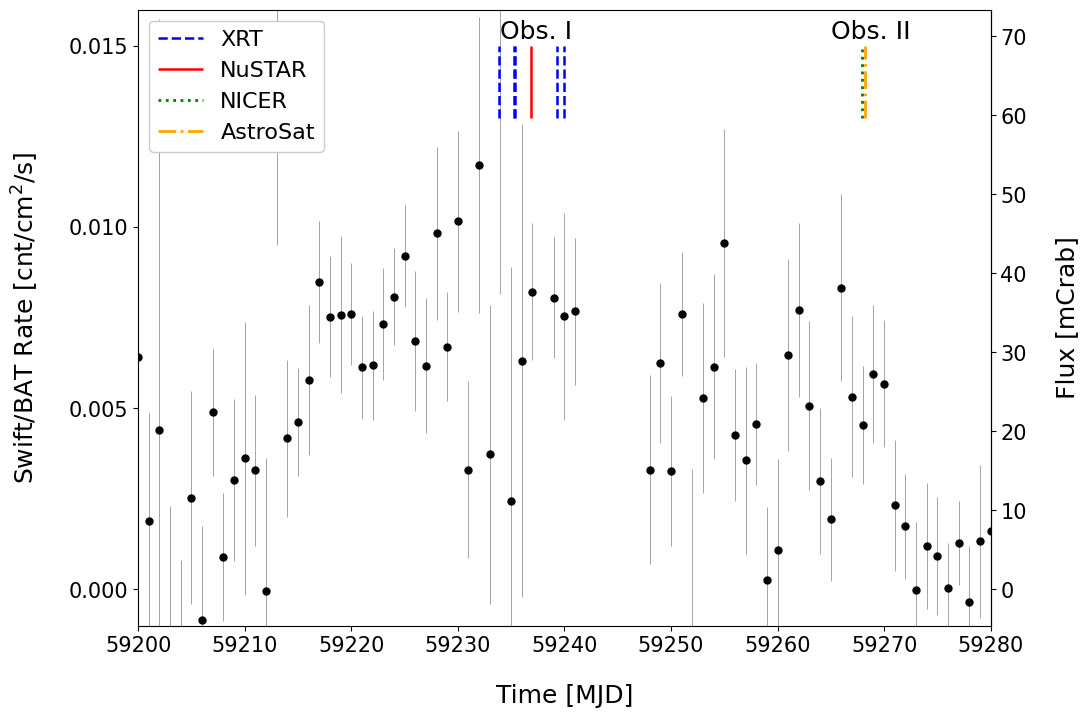}
\caption{\textsl{Swift}/BAT daily average light curve of 2S 1553-542 during the outburst in 2021 (black dots with grey error bars). Start times of each pointed observation are marked by vertical colored lines as detailed in the legend (same XRT ObsIDs may have slightly different MJDs). Separately analyzed observations (Obs. I and Obs. II) are also labelled as detailed in Sect.~\ref{sec:data_reduction} and Table~\ref{table:spectral}.\label{fig:outburst}}
\end{figure}

As observed by \textit{Chandra} and \textit{Swift}/XRT \citep[]{Lutovinov12} and \textit{RXTE}/PCA \citep[]{Pahari+Pal2012}, the source average spectrum can be fit with a highly absorbed (N$\rm _H\sim10^{22}\,$cm$^{-2}$) cutoff power-law modified by a blackbody component (kT$\sim2-4\,$keV) plus an Iron K$\alpha$ emission line at 6.4 keV. 
Moreover, during the previous outburst observed with \textit{NuSTAR} in 2015, a Cyclotron Resonant Scattering Feature (CRSF, or cyclotron line) was identified in the spectrum around 25 keV \citep{Tsygankov16}.
%Similarly to other spectral features, a cyclotron line is an important diagnostic that, when detected, helps pinning down some primary physical properties of the accreting NS, such as the magnetic field strength, as well as probing the accretion physics at different accretion regimes.
A cyclotron line is an important diagnostic of the physics in
the X-ray emission region. 
The energy of the fundamental line,
%In fact, the fundamental energy of a cyclotron line
$E_{\rm cyc}$, probes the magnetic field strength at the site of spectral emission, $E_\mathrm{\rm cyc}\sim11.6{\times}B_{12} (1 + z_\mathrm{g})^{-1}\,$keV, where $B_{12}$ is the magnetic field in units of $10^{12}\,$G, and $z_\mathrm{g}$ is the gravitational redshift (see \citealt{Staubert19} for a recent review).
Moreover, $E_{\rm cyc}$ has been observed to be luminosity-dependent in some sources \citep{Staubert19}. Such luminosity-dependence appears to be either positive or negative, according to the NS accretion regime. In fact, the dominant physical mechanisms that drive the accretion flow on the NS surface are distinguished according to the so-called critical luminosity, above which a radiation-dominated radiative shock occurs within the accreting structure at the magnetic poles  \citep{DavidsonOstriker1973,Basko+Sunyaev76,Becker+12, Mushtukov15_crit_lum}.

Recently, 2S 1553-542 has undergone a new outburst episode \citep{Nakajima19}, and we initiated a comprehensive observational campaign with \textit{NuSTAR, AstroSat} and \textit{NICER}, during which the source was also observed with \textit{Swift}/XRT.
Thanks to this, the source was observed at two different outburst stages -- one near the peak of the outburst, at a luminosity that is about $30\%$ brighter than that observed during the previous outburst, and the other towards the end of the outburst -- thus bracketing the luminosity previously covered by the NuSTAR observation in 2015.
Here we present the results of our study in terms of the spectral and timing characteristics of 2S 1553-542 and argue that the source has been observed undergoing an accretion regime transition.

\section{Data reduction}\label{sec:data_reduction}

\begin{table}[!t]
\caption{Log of the source observations used in this work.}
\label{table:log}
\centering
 \begin{tabular}{l c c c} 
 \hline
 & ObsID & MJD & Exposure \\
  & & (Start) & [ks] \\[0.5ex] 
 Swift/XRT & 000310960[08-10] & 59233.8 & 4.3 \\ 
 NuSTAR & 90701302002 & 59236.9 & 28.3 \\
 SXT & 9000004204 & 59268.2 & 18.8 \\
 LAXPC & 9000004204 & -- & 37.0 \\
 NICER & 3202030101 & 59268.1 & 1.4 \\ [1ex] 
 \hline
\end{tabular}
\end{table}

A log of all used observations is shown in Table~\ref{table:log}, while a light curve of the outburst is shown in Fig.~\ref{fig:outburst}. 
Hereafter we refer to Obs. I as that including \textit{Swift}/XRT and \textit{NuSTAR} data, and to Obs. II as that including \textit{NICER} and \textit{AstroSat} data, which were taken almost simultaneously, respectively.
Spectra were rebinned to have at least $50$ counts per bin.
As advised by the instruments teams, a systematic error of 1\% has been applied for all spectra from Obs.~II. 
Spectral data were analyzed using \texttt{XSPEC} v12.11.1c \citep{Arnaud96}.
\clearpage
\subsection{NuSTAR}
\textit{NuSTAR} \citep{Harrison13} was launched in 2012. It is currently the only X-ray mission with a telescope able to focus hard X-rays up to 79 keV.
\textit{NuSTAR} consists of two identical co-aligned telescopes that focus X-ray photons onto two independent Focal Plane Modules, FPMA and FPMB.
At the focus of each telescope module are four ($2\times2$) solid-state cadmium zinc telluride (CdZnTe) imaging detectors.
These provide wide-band (3--79\,keV) energy coverage with a FWHM of $18\arcsec$ and a spectral resolution of 400\,eV at 10\,keV. 

\textit{NuSTAR} observed 2S 1553-542 on 2021 January 22 (ObsID 90701302002, MJD 59236, \citealt{Malacaria_Atel21}).
The total exposure time was about 28 ks.
\textit{NuSTAR} data were reduced with \texttt{NUSTARDAS} v2.0.0 provided by the \texttt{HEASOFT} v6.28 and using the \textit{CALDB} 20210202 \citep{Madsen20}.
Cleaned events were obtained following the standard \textit{NuSTAR} guidelines.
Source spectra were extracted through the \texttt{NUPRODUCTS} routine.
The source extraction region was a $70\arcsec$ radius circular region centered on the source, while the background was extracted from a source-free region of comparable radius on the same detector (Det $0$) for FPMA and on the adjacent detector (Det $1$) for FPMB.
\textit{NuSTAR} spectral data were used in the range \mbox{$4-60\,$}keV (background counts dominate the spectrum above 60 keV).

%\textit{NuSTAR} source and background light curves were extracted with a 0.01 s bin size using the \texttt{nuproducts} tool. \textit{NuSTAR} light curves were barycentered using the \texttt{barycorr} tool and the \textit{NuSTAR} clock correction file \texttt{nuCclock20100101v118}.
%To ensure a better agreement between \textit{NuSTAR} and \textit{Swift}/XRT, we excluded \textit{NuSTAR} data below 4 keV.

\begin{figure}[!t]
\includegraphics[width=0.48\textwidth]{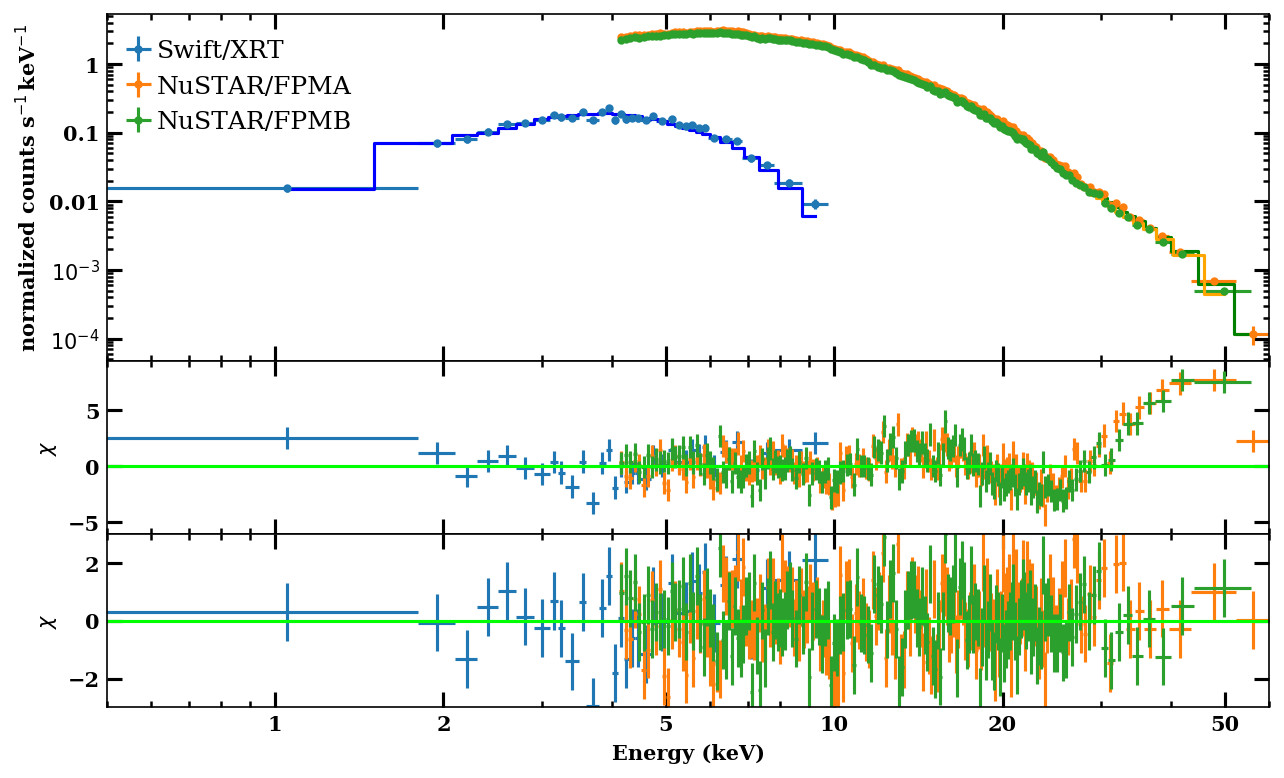}
\caption{\textit{Top}: 2S 1553-542 spectrum as observed by \textit{Swift}/XRT (blue) and \textit{NuSTAR} (FPMA and FPMB, orange and green, respectively) in 2021 and fit with a \texttt{cutoffpl} model (Table~\ref{table:spectral}, Obs. I).
\textit{Middle}: residuals of the \texttt{cutoffpl} model without a Gaussian absorption component.
\textit{Bottom}: residuals of the best-fit \texttt{cutoffpl} model including a Gaussian absorption line at $\sim27\,$keV (see Table~\ref{table:spectral}).
Spectra and residuals have been rebinned for plotting purpose.
%The blue text in the right corners of the lower panels shows the correspondent model $\chi^2$ divided by $\nu$ degrees of freedom.
\label{fig:spec1}}
\end{figure}

\subsection{Swift}

The \textit{Neil Gehrels Swift Observatory} \citep{Gehrels_2004} carries three scientific instruments covering a broad energy range of ${\sim}0.002-150\,$keV: the Burst Alert Telescope (BAT, $15-50$ keV), the X-ray Telescope (XRT, $0.5-10$ keV), and the UV/Optical Telescope (UVOT, not considered in this work).
\textit{Swift}/BAT performs daily scans of the X-ray sky \citep{Krimm13} and the daily light curve of 2S 1553-542 is here adopted from the public BAT transients monitor\footnote{\url{https://swift.gsfc.nasa.gov/results/transients/weak/H1553-542/}} (see Fig.~\ref{fig:outburst}).
On the other hand, \textit{Swift}/XRT performed pointed observations of 2S 1553-542 during the entire outburst episode with variable cadence, starting on 2020 January 7 (MJD 59221) through 2021 March 31 (MJD 59304).
However, to ensure similar physical conditions as those sampled by the \textit{NuSTAR} observation, we only analyzed adjacent \textit{Swift}/XRT ObsIDs, that is 00031096008 through 00031096010. We verified that the best-fit spectrum of each ObsID was in agreement with each other, and then merged the data to improve statistics.

\textit{Swift}/XRT data were reduced with \texttt{XRTDAS} v3.6.0, provided by the \texttt{HEASOFT} v6.28 and using the \textit{CALDB} 20200724.
Given the relatively high count rate of the source as detected by XRT ($\sim1.5\,$c/s) we excised the source core using a $13\arcsec$ extraction region in order to avoid pile-up effects.
The background was selected from an annular region centered on the source.
All \textit{Swift}/XRT observations were carried out in Photon Counting mode, whose timing resolution is insufficient for a timing analysis of the pulse profile for 2S 1553-542.

\begin{figure}[!t]
\includegraphics[width=0.48\textwidth]{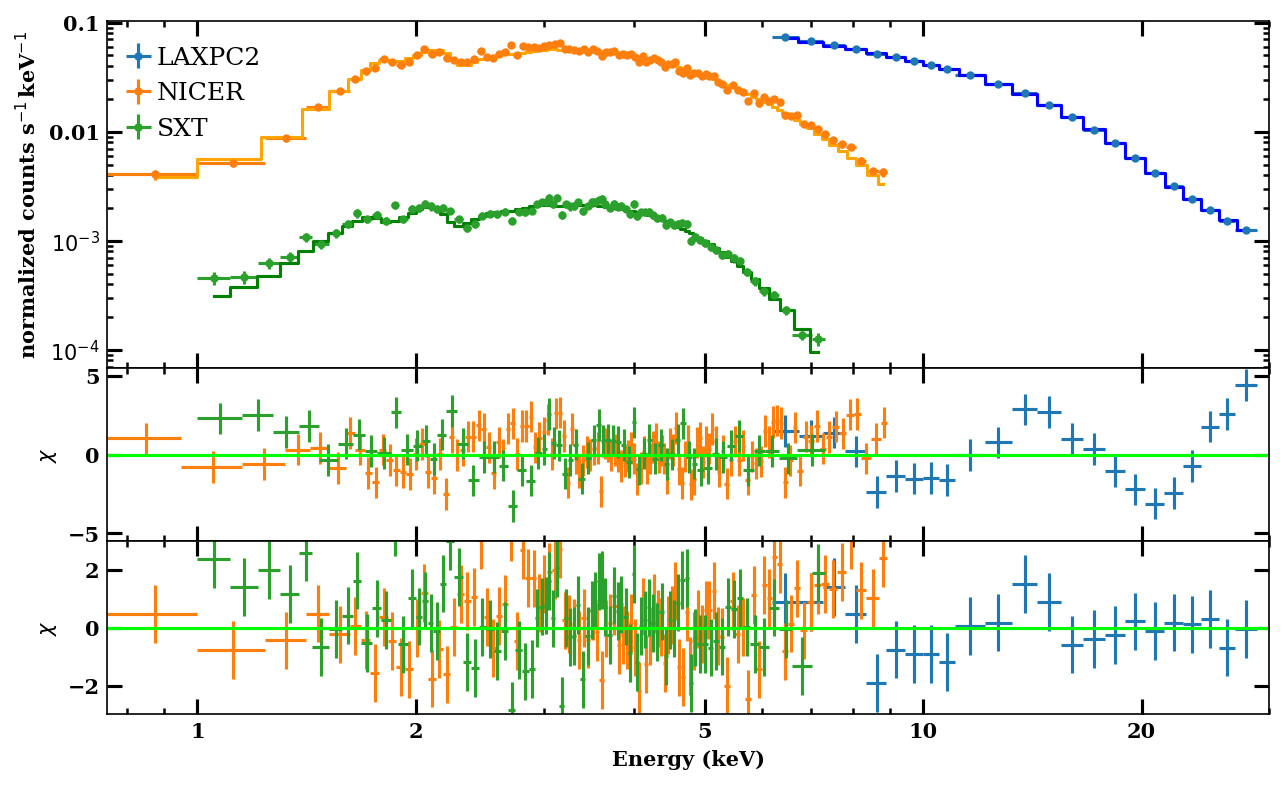}
\caption{
\textit{Top}: 2S 1553-542 spectrum as observed by \textit{NICER} (orange) and \textit{AstroSAT} (SXT and LAXPC2, green and blue, respectively) in 2021 and fit with a \texttt{cutoffpl} model (Table~\ref{table:spectral}, Obs. II).
\textit{Middle}: residuals of the \texttt{cutoffpl} model without a Gaussian absorption component.
\textit{Bottom}: residuals of the best-fit \texttt{cutoffpl} model including a Gaussian absorption line at $\sim23\,$keV (see Table~\ref{table:spectral}).
Spectra and residuals have been rebinned for plotting purpose.
\label{fig:spec2}}
\end{figure}

\clearpage
\subsection{AstroSat}

\textit{AstroSat} is the first Indian multi-wavelength space observatory. It was launched on September 28, 2015, and carries five scientific instruments on board \citep{Singh2014}. In this work we will use data only from the Soft X-ray focusing Telescope (SXT), and the Large Area X-ray Proportional Counter 2 (LAXPC20).
Following our Discretionary Director's Time request (PI: Malacaria), \textit{AstroSat} observed 2S 1553-542 on Feb. 23, 2021 (ObsID T03\_272T01\_9000004204).

\subsubsection{SXT}\label{subsubsec:sxt}

The SXT instrument on-board AstroSat is a  grazing incidence X-ray telescope capable to focus X-rays in the $0.3-8\,$keV nominal energy band, with a thermo-electrically cooled CCD detector in the focal plane. 
SXT observed 2S 1553-542 in Photon Counting (PC) mode with an exposure of $20\,$ks.
SXT data were processed following the official SXT walkthrough webpage\footnote{\url{http://astrosat-ssc.iucaa.in/uploads/sxt/SXT_walkthrough.pdf}} and using the \texttt{sxtpipeline v1.4b} software with the newest available CALDB (20160505).
Merged cleaned events were obtained with the SXT Event Merger Tool\footnote{\url{https://www.tifr.res.in/~astrosat_sxt/dataanalysis.html}.}.
In a fashion similar to \citet{Chaudhury2018}, we verified the optimal extraction region for the source events and selected a circular region with $10\,$arcmin radius centered at the source position corrected for the SXT misalignment\footnote{\url{http://astrosat-ssc.iucaa.in/uploads/APPS/NoteOnRelativeAngleBetwwenPayloads_Astrosat_15072016.pdf}}. 
As suggested by the official SXT instrument team\footnote{\url{https://www.tifr.res.in/~astrosat_sxt/dataanalysis.html}}, the employed background spectrum is the one distributed by the Tata Institute of Fundamental Research payload operation centres (TIFR-POC).
The employed Redistribution Matrix File (RMF) is made available from the SXT official team (\texttt{sxt\_pc\_mat\_g0to12.rmf}, version June 13, 2020), and includes the full grade $0-12$ range.
Using the \texttt{sxtARFModule} software, the standard SXT Ancillary Response File (ARF) was corrected  to account for the observation-specific parameters like the area of the source extraction region and vignetting effects.
Given the relatively low source count rate ($\sim2\,$cnt\,s$^{-1}$) no pile-up correction was necessary.
Finally, since the SXT data reduction pipeline does not correct for unaccounted gain drifts in the RMF, the \texttt{gain fit} tool in \texttt{XSPEC} was used to calculate the best-fit energy gain offset, while freezing the gain slope at $1$ \citep{Antia2021}.
This resulted in a positive gain offset of $0.048\pm0.009\,$keV (similar to \citealt{Chaudhury2018,Chakraborty20}), which was subsequently applied to the SXT RMF for optimal fitting.
Similarly to \citet{Bhargava19}, SXT spectral data are here only considered in the energy band $1-7\,$keV.

\subsubsection{LAXPC}

There are in total three identical LAXPC units on-board \textit{AstroSat}, namely LAXPC10, LAXPC20, and LAXPC30.
They are all Xenon+Methane, high-pressure (2 atm) proportional counters, and together cover an effective area of about $6000\,$cm$^2$ in the $5-20\,$keV energy band.
Each LAXPC detects X-ray photons in the energy range $3-80\,$keV with an energy resolution of $20\%$ at 30 keV \citep{Antia2021}.
However, LAXPC30 has been switched off in 2018 due to abnormal gain changes, while LAXPC10 is currently working at gain lower than nominal. Therefore, in this work only LAXPC20 data have been used. 
LAXPC data have been reduced with LAXPCSoftware\footnote{\url{http://astrosat-ssc.iucaa.in/laxpcData}} (Format A, version 20200804). % and with laxpcsoft (v3.3). 
For optimal S/N ratio, only events from the top layer have been considered.
The official \textit{AstroSat} tools released with the data reduction software were used to obtain source and background spectra in the $5-30\,$keV, as well as source and background light curves, following the method outlined in \citet{Misra2021}.
Similarly to SXT (see Sect.~\ref{subsubsec:sxt}), the best-fit energy gain offset has been calculated through the \texttt{gain fit} tool in \texttt{XSPEC} and resulted in an offset of $-0.52\pm0.05\,$keV (in agreement with \citealt{Antia2021}).
%Source and background spectra were obtained in the $5-30\,$keV band following the method outlined in \citet{Misra2021} using the official \textit{AstroSat} tools released with the data reduction software. Similarly for the source and background light curves.
%As a sanity check, a background spectrum was also estimated using Earth occultation times (see, e.g. ), and compared to that obtained from the high-energy background events.
The background spectrum was also rescaled for the deadtime effect.
We also verified that the LAXPC Field of View ($1^\circ\times1^\circ$) was free from significantly contaminating X-ray sources.

%LAXPC20 light curves were barycentered using the online orbit file generator\footnote{\url{http://astrosat-ssc.iucaa.in:8080/orbitgen/}.} and the official AstroSat tool \texttt{as1bary}\footnote{\url{http://astrosat-ssc.iucaa.in/?q=data_and_analysis}.}.

\subsection{NICER}

\textit{NICER} \citep{Gendreau2017} is an X-ray telescope deployed on the International Space Station (ISS) in 2017 June. \textit{NICER} X-ray Timing Instrument (XTI) has 56 aligned Focal Plane Modules (FPMs, 52 currently operational), each made up of an X-ray concentrator optic associated with a silicon drift detector. Together, all FPMs result in a peak collecting area of $1900\,$cm$^2$ at 1.5 keV. \textit{NICER} is capable of fast-timing observations in the $0.2-12.0$ keV band, with timing accuracy of time-tagged photons to better than 100 ns.
\citep{Prigozhin2016, LaMarr2016, Gendreau2017, Okajima2016}

Following our Discretionary Director's Time proposal, \textit{NICER} observed 2S 1553-542 in coordination with \textit{AstroSat} observations on Feb. 23, 2021 for a total exposure time of 1.4 ks (ObsID 3202030101), and successively on Feb. 26 (ObsID 3202030102), but only data from the former observation have been considered in this work (given its contemporaneity with \textit{AstroSat} observations).
NICER data were processed with HEASoft version 6.28 and the NICER Data Analysis Software  (\texttt{nicerdas}) version 7.0 (\texttt{2020-04-23V007a}) with Calibration Database (CALDB) version \texttt{xti20200722}, adopting standard calibration and screening criteria implied in the \texttt{nicerl2} tool.
The background spectrum was obtained using the space-weather method implemented in the version \texttt{v0p6} of the \texttt{nicer\_bkg\_estimator}\footnote{\url{https://heasarc.gsfc.nasa.gov/docs/nicer/tools/nicer_bkg_est_tools.html}} tool (Gendreau et al., in prep.).
The energy band of \textit{NICER} spectrum was limited to $0.75-9\,$keV below which the spectrum was background dominated and to avoid potential noise at the high-energy end of the bandpass.

\section{Results}
\subsection{Spectral analysis}\label{subsec:spectral}

In order to compare our results with those available from \citet{Tsygankov16}, we adopted a similar spectral model, that is an absorbed cutoff power-law model (\texttt{cutoffpl} in \texttt{XSPEC}) modified with a soft blackbody (\texttt{bbodyrad}), a Fe K$\alpha$ emission line (\texttt{gauss}) at 6.4 keV, and an absorption line with Gaussian optical depth profile (\texttt{gabs}) to take into account the CRSF.
Different continuum components (\texttt{Highecut, CompTT}) were tested in place of the cutoff power-law component and they all require a cyclotron line to fit the data.
The photoelectric absorption component \texttt{tbabs} from \texttt{XSPEC} was used assuming model-relative (\texttt{wilm}) elemental aundances \citep[]{Wilms00}.
Moreover, since the Obs. II is carried out at a luminosity level that is lower than that analyzed in \citet{Tsygankov16}, a few additional measures need to be taken.
In particular, the spectrum obtained from Obs.~II does not require a Fe K$\alpha$ emission line, while it requires a second, colder blackbody component.
 
\begin{table}[!t]
\caption{Best-fit results of 2S 1553-542 spectral analysis with a cutoff power-law model \texttt{cutoffpl} combined with one or two blackbody components, an Iron K$\alpha$ line and a cyclotron line. All reported errors are at $1\sigma\,$c.l., obtained using the \texttt{err} tool from \texttt{XSPEC}.} \label{table:spectral}
\begin{ruledtabular}\begin{tabular}{lcc}
 & Obs I & Obs II \\
  & \\
C$_{\rm FPMA}$ &  1 (fixed) & -- \\
C$_{\rm FPMB}$ & $1.045^{+0.002}_{-0.002}$ & -- \\
C$_{\rm XRT}$ & $1.16^{+0.02}_{-0.02}$ & -- \\
C$_{\rm LAXPC}$ & -- & 1 (fixed) \\
C$_{\rm NICER}$ & -- & $0.98^{+0.01}_{-0.01}$ \\
C$_{\rm SXT}$ & -- & $1.05^{+0.01}_{-0.02}$ \\
N$_{\textrm{H}}$ [$10^{22}\,$cm$^{-2}$] & $4.7^{+0.3}_{-0.4}$ & $4.1^{+0.2}_{-0.5}$ \\
kT$_{\rm Cold BB}\,$[keV] & -- & $0.049^{+0.003}_{-0.003}$ \\
norm$_{\rm Cold BB}$ & -- & $(1.4^{+3.6}_{-1.2})\times10^{11}$ \\
kT$_{\rm HotBB}\,$[keV] & $1.29^{+0.05}_{-0.02}$ & $1.68^{+0.07}_{-0.06}$ \\
norm$_{\rm Hot\,BB}$ & $10.2^{+1.3}_{-0.9}$  & $1.3^{+0.2}_{-0.2}$ \\
E$_{\rm K\alpha}\,$[keV] & $6.45^{+0.05}_{-0.04}$ & -- \\
$\sigma_{\rm K\alpha}\,$[keV] & $0.30^{+0.06}_{-0.06}$ & -- \\
norm$_{\rm K\alpha}$ [ph/cm$^2$/s] & ($4.4^{+0.7}_{-0.8})\times10^{-4}$ & -- \\
$\Gamma$ & $-0.7^{+0.3}_{-0.4}$ & $0.29^{+0.07}_{-0.06}$ \\
HighECut [keV] & $5.6^{+0.4}_{-0.5}$ & $9.9^{+1.4}_{-0.8}$ \\
norm$_{\Gamma}^*$ & $0.007^{+0.004}_{-0.003}$ & $0.013^{+0.001}_{-0.001}$ \\
E$_{\rm cyc}\,$[keV] & $27.2^{+0.3}_{-0.3}$ & $22.9^{+0.8}_{-0.6}$ \\
$\sigma_{\rm cyc}\,$[keV] & $6.9^{+0.5}_{-0.5}$ & $3.1^{+0.7}_{-0.6}$ \\
Strength$_{\rm cyc}$ [keV] & $11.2^{+2.2}_{-1.6}$ & $2.3^{+1.3}_{-0.7}$ \\
Flux$^\dagger$ (3-30 keV)& $1.481^{+0.003}_{-0.003}\times10^{-9}$ & $7.34^{+0.03}_{-0.03}\times10^{-10}$ \\
Flux$^\dagger$ (3-20 keV)& $1.321^{+0.002}_{-0.002}\times10^{-9}$ & $6.09^{+0.03}_{-0.03}\times10^{-10}$ \\
$\chi^2$/d.o.f. & $1338/1223$ & $887/791$ \\
\end{tabular}\end{ruledtabular}
\tablenotetext{}{
$^*$In units of photons/keV/cm$^2$/s at 1 keV.\quad$^\dagger$Flux calculated for the entire model and reported in units of erg\,cm$^{-2}\,$s$^{-1}$. Flux values with estimated errors were obtained using the \texttt{cflux} model from \texttt{XSPEC} as resulting from FPMA and LAXPC in Obs. I and II, respectively.}
\end{table}
 
%A systematic error of 1\% has been applied for the spectra from Obs.~II. 
A cross-normalization constant was applied to take into account uncertainties in calibration among the various instruments.
The seemingly large cross-normalization value between \textit{Swift}/XRT and \textit{NuSTAR}, $\sim16\%$, is in fact consistent with the expected range from \citet{Madsen15} and others (see, e.g., \citealt{Molina19}), and also reflects the intrinsic source variability between observations (see Fig.~\ref{fig:outburst}).
%For the Iron K$\alpha$ line in Obs.~II we set upper limits by including in the model a Gaussian line with same energy and width as in Obs.~I, and letting the normalization free to vary. The best fit does not significantly improve the $\Chi^2$, but returns a normalization value of $1.7^{+1.7}_{-1.6}\times10^{-4} ph/cm2/s$ (errors at 90\% c.l.).
%These values can be converted in Equivalent Width by using eqwidth [eqwidth #component] tool in XSPEC, which returns $0.034^{+0.034}_{-0.033}$ keV.
% Using 6.45 keV = 1.033x10^-8 ergs, we obtain upper limit:
%1.033x10^-8 ergs * 1.7x10^-4 ph/cm2/s = 1.8^{1.8}_{-1.6} erg/cm^2/s.

\begin{figure}[!t]
\includegraphics[width=0.48\textwidth]{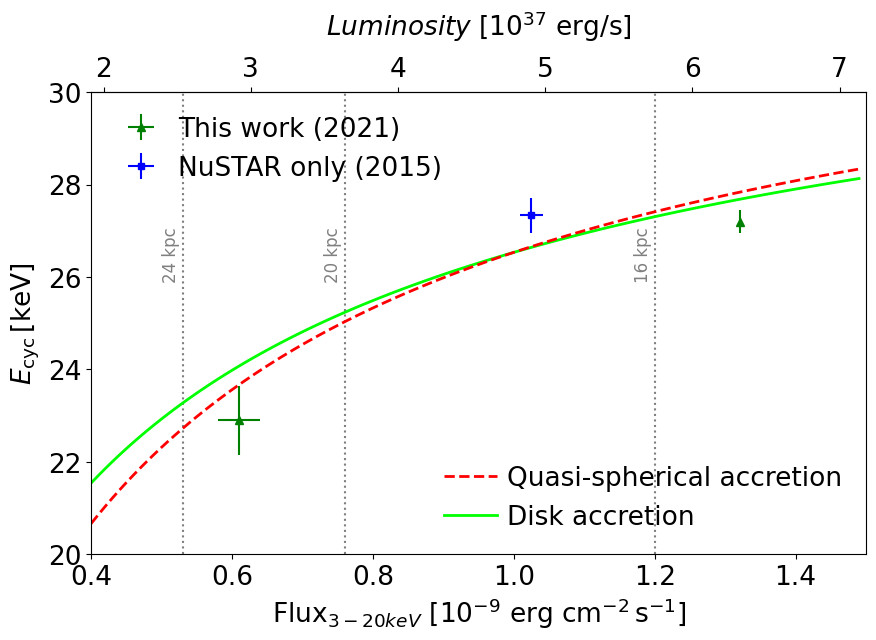}
\caption{Cyclotron line energy E$_{cyc}$ as a function of flux (bottom x-axis) and luminosity (top x-axis, d${=}20\,$kpc) in the 3-20 keV energy range. Error bars indicate the $1\sigma$ c.l. 
Lime continuous line and red dashed line show a fit to the data points with the collisionless shock model (see Sect.~\ref{subsec:spectral_discussion}) for the disk accretion and the quasi-spherical settling accretion, respectively.
%disk accretion and quasi-spherical settling accretion models, respectively (see Sect.~\ref{subsec:spectral_discussion}). 
Grey vertical lines show the critical luminosity $L_{crit}$ (from Eq.~\ref{eq:BW}) obtained from the bottom x-axis fluxes for different
distance values (d${=}20\pm4\,$kpc).
\label{fig:correlation}}
\end{figure}

Our spectral analysis from two different outburst stages is presented in Table~\ref{table:spectral} and shows a variation of the cyclotron line energy with luminosity (see Fig~\ref{fig:correlation}).
A comparison of our results with those obtained from the \textit{NuSTAR} observation of 2S 1553-542 in 2015 \citep[Model II solution in their work]{Tsygankov16} is also shown in Fig~\ref{fig:correlation}.
%A linear fit to the 2021 outburst data returns a slope $m=6.05\pm$ and an intercept $b=$, while the linear fit to the 2021 data point at lowest flux and the data point from 2015 returns a slope $m=\pm$ and an intercept $b=$.
For reference, the best-fit Model II in \citet{Tsygankov16} shows a spectral photon index $\Gamma=-0.66$, a cutoff energy E$_{\rm cut}=5.1$\,keV, a blackbody component with kT$_{\rm BB}=0.94$ and norm$_{\rm BB}\simeq20$, an Iron K$\alpha$ line at $6.45\,$keV and $\sigma_{\rm K\alpha}=0.4\,$keV, and a cyclotron line with E$_{\rm cyc}=27.3\,$keV, $\sigma_{\rm cyc}=6.4\,$keV, $\tau_{\rm cyc}=8.3\,$keV.
We also notice that, similarly to \citet[]{Tsygankov16}, an uncommonly hard photon index was obtained for the Obs. I spectrum (see Table~\ref{table:spectral}), in opposition with the Obs. II spectral photon index.
We therefore verified that the best-fit solution was not resulting from the lack of a cold blackbody component in Obs. I which, given the large emitting radius, may represent the emerging radiation from the accretion disk becoming relevant as the main source continuum becomes fainter. 
Restricting the photon index to positive values only and adding a cold blackbody component (best-fit values kT$_{\rm Cold\,BB}\sim0.18$, norm$_{\rm Cold\,BB}\sim2.8E4$) results in a slightly worse $\chi^2/d.o.f.=1346/1221$. %=1.102$,
%(against the $\chi^2/d.o.f.=1338/1223=1.092$), 
In this case, other model parameters remain roughly consistent with those reported in Table~\ref{table:spectral}, with the largest difference, i.e. $\sim40\%$, shown by the normalization of the hot blackbody component, norm$_{\rm Hot\,BB}=7.3_{-0.9}^{+1.2}$, while the best-fit value of the photon index, $\Gamma=0.0^{+0.4}_{-0.0}$, gets pegged at its lower limit. 
Even in the case where the photon index $\Gamma\geq0$ and a cold blackbody component is included, the Obs. I best-fit model returns a value of the cyclotron line energy that remains almost unchanged, E$_{\rm cyc} = 27.1_{-0.3}^{+0.3}\,$keV.

\begin{figure}[!t]
\includegraphics[width=0.48\textwidth]{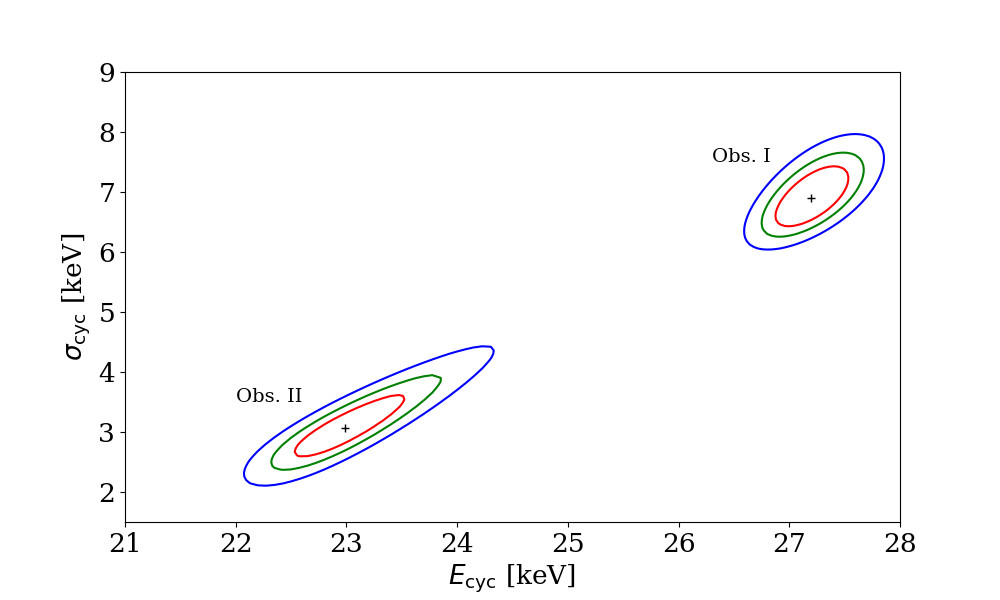}
\caption{$\chi^2$-contour plots for two parameters, E$_{\rm cyc}$ and $\sigma_{\rm cyc}$ from Obs. I and II, as indicated.  Red, green and blue contours correspond to the $68\%$-uncertainty for one and two parameters of interest and to the $90\%$-uncertainty for two parameters of interest, respectively. The black cross in the midst of the contours indicates the best-fit solution.
\label{fig:contours}}
\end{figure}

To rule out possible artificial (model-driven) dependencies of the cyclotron line energy E$_{\rm cyc}$ and other model parameters, we analyzed the correspondent $\chi^2$-contour plots. 
No significant dependence of the E$_{\rm cyc}$ on any continuum parameters was found, but the CRSF Gaussian parameters were found to be correlated one another. 
However, confidence level contour plots clearly show separated intervals, thus supporting the physical interpretation of the cyclotron line energy luminosity-dependence. This is shown, for example, in Fig.~\ref{fig:contours} for E$_{\rm cyc}$ versus $\sigma_{\rm cyc}$ (with a similar behaviour for E$_{\rm cyc}$ versus Strength$_{\rm cyc}$ contour plots). 

\subsection{Timing analysis}

To more thoroughly investigate possible changes in the accretion regime during our observations, we extracted energy-dependent pulse profiles from the source at different luminosities.
To do so, we applied the following corrections.

For \textit{NuSTAR}, source and background light curves were extracted with a 0.1 s bin size using the \texttt{nuproducts} tool. \textit{NuSTAR} light curves were barycentered using the \texttt{barycorr} tool and the \textit{NuSTAR} clock correction file \texttt{nuCclock20100101v118}.
LAXPC20 
%events and GTIs were ... and 
light curves also were extracted with a 0.05 s bin size and barycentered using the online orbit file generator\footnote{\url{http://astrosat-ssc.iucaa.in:8080/orbitgen/}.} and the official AstroSat tools for Header keywords correction \texttt{prepbary\_laxpc} and for Barycentric correction \texttt{as1bary}\footnote{\url{http://astrosat-ssc.iucaa.in/data_and_analysis}.}.
The same corrections (except the background subtraction) were applied to the NICER light curve.
Orbital demodulation was also applied to all datasets, based on the recently updated orbital solution from \citet{Malacaria20}.
Pulse periods were obtained from the \textit{NuSTAR} FPMA and FPMB combined, background-corrected light curves for Obs. I, and from \textit{AstroSat}/LAXPC20 background-corrected light curves for Obs. II.
Pulse periods were determined using the epoch folding technique \citep{Leahy+83} through the \texttt{efsearch} HEASARC tool.
The process results in a pulse period value of 9.282155(3) s for Obs. I and 9.279490(8) for Obs. II. 
The spin period derivative is therefore $\dot{P\rm _s}=-9.8\times10^{-10}\,$s\,s$^{-1}$ ($\dot{\nu}=1.1\times10^{-11}\,$Hz\,s$^{-1}$).
Pulse periods uncertainties were estimated by simulating light curves from the previously determined pulse profiles, following the method outlined in \citet[]{Lutovinov12, Boldin13}. $10^3$ light curves were simulated, with count rates generated randomly within the error of the original data. Epoch folding was then applied to each simulated light curve to obtain a pulse periods distribution. The standard deviation of such distribution was taken as the pulse period uncertainty of the data.

To allow a comparison with the pulse-profiles obtained by \citet{Tsygankov16}, we extracted pulse profiles in the same energy bands, i.e., 3-7, 7–18, 18–30 keV for our \textit{NuSTAR} and \textit{AstroSat}/LAXPC observations. %(plus the additional 30–50 keV band for \textit{NuSTAR}).
The \textit{NICER} pulse profile in the 1-3 keV energy band was also extracted.
Energy- and luminosity-resolved pulse profiles are shown in Fig.~\ref{fig:pulse_prof}.
Due to the non-monotonic variation of the pulse period throughout the outburst and to the residual wave-like behavior of the spin frequency evolution around the best orbital solution seen in \textit{Fermi}/GBM\footnote{\url{https://gammaray.nsstc.nasa.gov/gbm/science/pulsars/lightcurves/2s1553.html}}, pulse profiles at different times are not phase connected. 
Therefore, in Fig.~\ref{fig:pulse_prof} they have been manually aligned with respect to the peak of the \textit{NuSTAR} pulse profile in the 7-18 keV energy band.

\section{Discussion}

\subsection{Spectral analysis}\label{subsec:spectral_discussion}
A luminosity-dependence of the parameters characterizing the XRPs spectral model is observed in many sources and expected from theoretical arguments (see, e.g., \citealt{Staubert19}, and references therein).
According to theoretical models \citep{Basko+Sunyaev76, Becker+12, Poutanen13,Mushtukov15_crit_lum, Mushtukov15_pos_corr}, two different accretion regimes are at work in the accretion structure, depending on the accretion rate, and therefore luminosity.
These two different accretion regimes are separated by a critical luminosity, $L_{\rm crit}$, characterized by a radiation pressure so intense that prevents the accreting matter to free-fall on the NS surface, giving rise to a radiative shock and forming an accretion column.
%Depending on the physical parameters of the NS and its accreting modality (i.e., disk- or  wind-dominated accretion), the typical value of $L_{crit}$ is of the order of $10^{37}\,$erg/s.
For a value of the cyclotron line energy of E$_{\rm cyc}=27\,$keV, the critical luminosity obtained by \citet[see their Eq.~55]{Becker+12} is:
\begin{equation}\label{eq:BW}
\begin{split}    
    L^a_{\rm crit} =& 1.28\times10^{37}\,erg\,s^{-1} \left(\frac{\Lambda}{0.1}\right)^{-7/5} \left(\frac{M_{\rm NS}}{1.4M_\odot}\right)^{29/30}\\& \left(\frac{R_{\rm NS}}{10\,km}\right)^{1/10} \left(\frac{E_{\rm cyc}}{10\,\rm keV}\right)^{16/15}
\end{split}
\end{equation}
and it is equal to $3.7\times10^{37}\,$erg\,s$^{-1}$ for $\Lambda=0.1$ the accretion flow geometry constant for disk accretion, M$_{\rm NS}=1.4\,M_{\odot}$ the mass of the NS, R$_{\rm NS}=12\,$km the radius of the NS.
On the other hand, for the same cyclotron line energy \citet{Mushtukov15_crit_lum} predicts a critical luminosity value $L_{\rm crit}^b\sim1\times10^{37}\,$erg/s  (see, e.g., their Fig.~7 for the case of pure X-mode polarization).
$L^a_{\rm crit}$ has been shown in Fig.~\ref{fig:correlation} for the upper, nominal and lower values of the source distance, $d=20\pm4\,$kpc.

\begin{figure}[!t]
\includegraphics[width=0.48\textwidth]{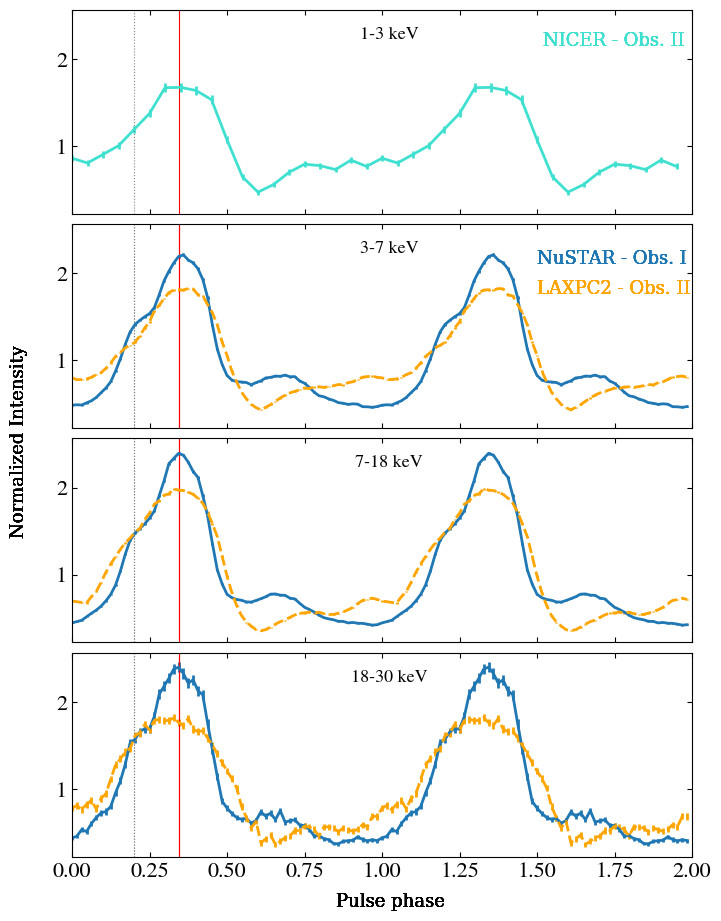}
\caption{
Energy- and luminosity-resolved pulse profiles of 2S 1553-542 as observed during the outburst in 2021 with \textit{NICER} (top panel), \textit{NuSTAR} and \textit{AstroSat} (remaining panels) in the energy bands 1-3, 3-7, 7-18, 18-30 keV from top to bottom. Pulse profiles are normalized by the average source intensity in each band. The vertical red line marks the main peak in the \textit{NuSTAR} (7-18 keV) pulse profile, with respect to which the other pulse profiles have been aligned.
The grey dotted line at $\phi\sim0.2$ marks the energy-dependent wing.
\label{fig:pulse_prof}}
\end{figure}

From an observational perspective (see, e.g.,  \citealt{Klochkov+11}, and references therein), some accreting XRPs exhibit a positive correlation between the E$_{\rm cyc}$ and luminosity when the source is accreting in the sub-critical ($L<L_{\rm crit}$) regime, while a few accreting XRPs show the opposite correlation in the super-critical ($L>L_{\rm crit}$) regime. 
On the other hand, some sources do not show any correlation (e.g., 4U~1538-522, \citealt{Hemphill16}), while a few sources show either both correlations (V~0332+53, \citealt{Doroshenko+17}, and GRO~J1008-57, \citealt{Chen21}), or a flattening of the parameters dependence \citep{Rothschild17, Vybornov17}.
For V~0332+53, the reported inversion of the cyclotron line energy luminosity-dependence has been interpreted as evidence for accretion regime transition. An analogous interpretation was suggested for GRO~J1008-57 based on the continuum spectral evolution solely, yet not supported by the pulse profile behaviour \citep{Kuehnel+13}.
%Theoretical modelling of the observed correlations at different accretion regimes are given in \citet[]{Becker+12, Poutanen13, Mushtukov15_pos_corr} and all interpret the correlations as a result of the physical changes in the accretion structure as the accretion rate varies.

Despite only a handful of data points being available in Fig.~\ref{fig:correlation}, the trend suggests a positive correlation of $E_{cyc}$ with luminosity and perhaps following flattening, similar to what has been observed in GX 304-1 \citep{Rothschild17} and Cep X-4 \citep[]{Vybornov17}.
However, the flattening in those two sources was observed at a sub-critical accretion regime, where the accretion flow possibly decelerates in  a collisionless shock \citep[]{Langer82} that only exists when the accretion rate is low (i.e., a few percent of the Eddington luminosity, \citealt{Braun84, Bykov04}).
For a typical NS mass $M_{\rm NS}=1.4\,M_\odot$, the Eddington luminosity is $L_{\rm Edd}=1.3\times10^{38}\,$erg\,s$^{-1}$, that is about $50\%$ larger than the highest observed luminosity value for 2S 1553-542 in this work.
Therefore, the collisionless shock should not play a dominant role at the accretion rates probed in this work, and the cyclotron line energy dependence on luminosity is not expected to flatten similarly to GX 304-1 and Cep X-4.
Nonetheless, given the trend observed in Fig.~\ref{fig:correlation} we tested the collisionless shock model by fitting the following function (see Eq.~6 from \citealt{Rothschild17}):
\begin{equation}
    E_{\rm cyc} (F\rm _x) = E_0 (K_1 F\rm _x^{-\alpha}+1)^{-3}
\end{equation}
where $E\rm _0$ is the cyclotron line energy resulting from the NS surface, $K_1 = H_{\rm CRSF}/R_{\rm NS}$ (with $H_{\rm CRSF}$ being the height within the accretion structure where the cyclotron line is formed) is assumed constant, $F\rm _x$ is the observed X-ray flux, and $\alpha=5/7$ for disk accretion. 
The fit returns $E_0=33.8\pm1.7\,$keV and $K_1=0.08\pm0.01$ ($10^{-9}$ erg cm$^{-2}$ s$^{-1}$)$^\alpha$.
A similar fit and best-fit values are obtained if $\alpha=9/11$ for quasi-spherical settling accretion is assumed.
These values are comparable to those obtained for GX~304-1 and Cep~X-4 \citep{Rothschild17,Vybornov17}.

%Say something more on Fig. 4 and how there could be a transition around the line at 20kpc.
On the other hand, the trend in Fig.~\ref{fig:correlation} may also be interpreted as indicative of an accretion regime transition around the $L_{\rm crit}$. 
The critical luminosity values shown in Fig.~\ref{fig:correlation} are derived from the observed flux assuming isotropic emission and their maximum uncertainty is estimated as $25\%$ (see \citealt[]{MartinezNunez17}).
In this interpretation, $E_{\rm cyc}$ correlates with luminosity until $L_{\rm crit}$ is reached, and then either flattens or decreases with luminosity. In this scenario, the distance value plays a key role. In fact, this scenario favors distance values $<20\,$kpc, with the accretion regime transition occurring at $L_{\rm crit}\approx4\times10^{37}\,$erg\,s$^{-1}$ obtained from Eq.~\ref{eq:BW}, corresponding to about $16\,$kpc in Fig.~\ref{fig:correlation}.
Further analysis of the pulse profiles and of the pulsed fraction
support this interpretation (see Sect.~\ref{subsec:timing_discussion}).
%will furnish additional evidence that the accretion regime transition happens at a critical luminosity corresponding to a distance of about 16 kpc in Fig.~\ref{fig:correlation} 

\subsection{Timing analysis}\label{subsec:timing_discussion}

Comparing the luminosity-dependence of the pulse profiles at different energy bands can provide insights on the accretion regime at work on the NS.
\citet{Tsygankov16} showed that the pulse profile of 2S 1553-542 is only moderately energy-dependent, with a main broad peak spanning roughly half pulse cycle, plus a trailing hard-energy wing that contributes to the main peak and that only shows up in the $18-30\,$keV energy band (i.e., the energy band sampling cyclotron line).
A similar trailing wing was observed in other sources as well \citep[]{Tsygankov06, Iyer15}.
Its interpretation is that of a phase-lagging feature around the cyclotron line energy due to the energy-dependent beaming of the radiation at the site of emission, with the emission around the cyclotron line energy escaping through a pencil beam while the remaining spectrum is emitted mostly in a fan beam fashion \citep[]{Ferrigno11, Schonherr14}.

\begin{figure}[!t]
\includegraphics[width=0.48\textwidth]{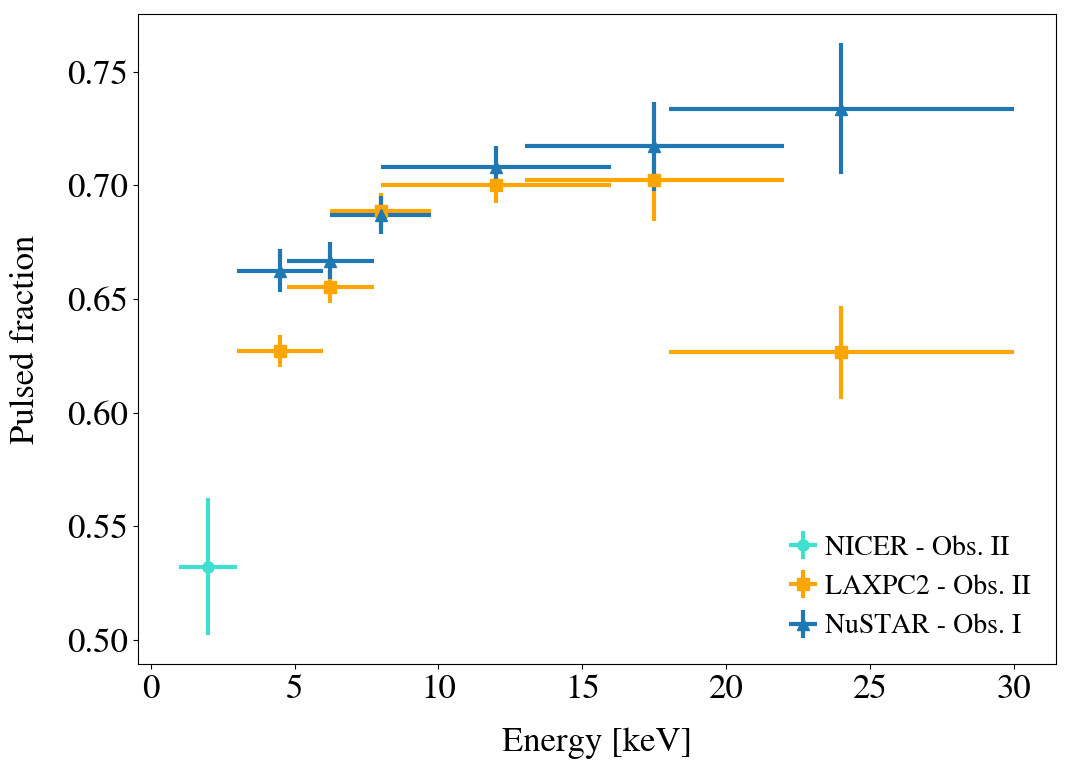}
\caption{Pulsed-fraction plotted as a function of the energy for the pulse profiles shown in Fig.~\ref{fig:pulse_prof}, subdivided in not independent energy-bins for better visualization. Cyan circles, blue triangles and orange squares data points correspond to  \textit{NICER}, \textit{NuSTAR} and \textit{AstroSat} pulse profiles, respectively.
\label{fig:pulsed_fraction}}
\end{figure}

The pulse profiles from our monitoring campaign (see Fig.~\ref{fig:pulse_prof}) show a similarly moderate energy-dependence as \citet{Tsygankov16}, with the exception that the hard-energy wing is present at all energy bands probed with our \textit{NuSTAR} observation (carried out at a source luminosity about $30\%$ higher).
The hard-energy wing only shows energy-dependence at the lower luminosity level of the \textit{AstroSat} observation (Obs. II).
Moreover, the \textit{AstroSat} pulse profiles morphology resembles that of the pulse profiles obtained with \textit{NuSTAR} by \citet{Tsygankov16} so strongly that they can be considered as representative of the same accretion regime.
In addition, a comparison of the luminosity-dependent profiles obtained in this work suggests a narrow beaming component emerging at higher luminosity (the main \textit{NuSTAR} peak at pulse phase $\phi\sim0.35$ in Fig.~\ref{fig:pulse_prof}).
The \textit{NuSTAR} profiles also show a secondary peak located around the main minimum of the \textit{AstroSat}/LAXPC profiles ($\phi\sim0.65$).
Finally, \textit{NICER} pulse profile resembles those from \textit{AstroSat}/LAXPC, coherently with the finding that pulse profiles from this source show only moderate energy-dependence but, at the same time, a rather marked luminosity dependence, consistent with an accretion regime transition once the critical luminosity $L_{\rm crit}\approx4\times10^{37}\,$erg/s is crossed.

To study the pulsed fraction (PF), we define it as ($I_{\rm max}-I_{\rm min}$)/($I_{\rm max}+I_{\rm min}$), where $I_{\rm max}$, $I_{\rm min}$ are the maximum and minimum pulse profile count rate, respectively.
The PF is high ($>50\%$) and shows a marked luminosity- and energy-dependence (see Fig.~\ref{fig:pulsed_fraction}).
At higher luminosity  (i.e., that covered by the \textit{NuSTAR} observation) the pulsed fraction keeps increasing with energy. However, at lower luminosity (i.e., that covered by the \textit{AstroSat} observation) the pulsed fraction shows a reversal trend around the cyclotron line energy, almost identical to that observed by \citet[]{Tsygankov16}, thus strengthening the interpretation that both observations are representative of the same, sub-critical accretion regime.
Local features around the cyclotron line energy in the energy-dependent pulsed fractions have been observed in other sources as well (see, e.g., \citealt[]{Ferrigno09}), and are usually ascribed to the resonant scattering that causes beamed emission to become isotropic.
Moreover, the stark difference in trend and value of the pulsed fraction in the 18-30 keV energy band seems to reflect the luminosity-dependence of the hard-energy wing observed in the pulse profiles.

Both the pulse profile and the pulsed fraction luminosity-dependence are generally attributed to the switch of the accretion column dominant beaming pattern, with the pencil beam dominating at lower luminosity and a fan beam growing contribution as the luminosity increases, with some hybrid configurations in between (see, e.g., \citealt[for theoretical models]{Basko+Sunyaev75, Basko+Sunyaev76, Blum+Kraus2000, Becker+12} and, e.g., \citealt[for observational results]{Wilson-Hodge18, Epili+17, Malacaria+15}).
In this scenario, the pulsed fraction switch in the 18-30 keV energy band proves additional evidence that the accretion regime has drastically changed as the luminosity increased and eventually crossed the critical luminosity value derived in Sect.~\ref{subsec:spectral_discussion}, $L_{\rm crit}\approx4\times10^{37}\,$erg/s.

%Typically, single-peaked pulse profiles are observed below the critical luminosity and are interpreted as originating from the hot spot at the base of the accretion mound in the form of a pencil beam, while more complex, multi-peaked pulse profiles emerge as the luminosity increases towards and above the critical luminosity and the fan beam component from the walls of the accretion column becomes more important (see, e.g., \citealt[]{Wilson-Hodge18, Epili+17, Malacaria+15}).

\section{Summary and conclusions}

We have analyzed the most recent outburst from the accreting XRP 2S 1553-542 at the beginning of 2021, taking advantage of a multi-observatory campaign. We performed spectral and timing analysis, and focused on the spectral and pulse profiles luminosity-dependence.

The spectral analysis reveals a positive correlation between the cyclotron line energy and the observed luminosity (Fig.s~\ref{fig:correlation}, \ref{fig:contours}), observed for the first time in this source. The correlation ceases above a certain luminosity, which can be interpreted either in terms of a collisionless shock, or as an inversion of the correlation above the critical luminosity $L_{\rm crit}$. 
The latter scenario is favored by additional evidence and supports a distance value for 2S 1553-542 of about 16 kpc, which is skewed towards the lower limit of the nominal distance value for this source.

The pulse profiles analysis also reveals a drastic change in the pulse profile shape once the source crosses the critical luminosity (Fig.~\ref{fig:pulse_prof}). This result is supported by a trend inversion of the pulsed fraction corresponding to those profiles, in an energy band that samples the cyclotron line (Fig.~\ref{fig:pulsed_fraction}).

Based on the above indications we conclude that we have witnessed an accretion regime transition, happening at a $L_{\rm crit}\approx4\times10^{37}\,$erg/s. Given the elusive nature of such phenomenon, the accretion regime transition has been rarely observed, making therefore 2S 1553-542 a key addition to the short list of XRPs whose accretion regime transition is supported by the luminosity-dependence of the cyclotron line energy, the pulse profiles and the pulsed fraction altogether.
%Future X-ray observations of 2S 1553-542 are particularly encouraged in the luminosity range corresponding to the critical luminosity, in order to obtain a denser sampling of the source spectral and timing properties at a luminosity stage where the most drastic changes are expected.
\acknowledgments
\balance

This research has made use of data and software provided by the High Energy Astrophysics Science Archive Research Center (HEASARC), which is a service of the Astrophysics Science Division at NASA/GSFC and the High Energy Astrophysics Division of the Smithsonian Astrophysical Observatory. We acknowledge extensive use of the NASA Abstract Database Service (ADS). The material is based
upon work supported by NASA under award number 80GSFC21M0002 (CRESST II). 
This work used data from the NuSTAR mission, a project led by the California Institute of Technology, managed by the Jet Propulsion Laboratory, and funded by NASA, and has utilized the NUSTARDAS software package, jointly developed by the ASI Science Data Center, Italy, and the California Institute of Technology, USA.
This work made use of data supplied by the UK Swift Science Data Centre at the University of Leicester and has also made use of the XRT Data Analysis Software (XRTDAS) developed under the responsibility of the ASI Science Data Center, Italy.
This work was supported by NASA through the NICER mission and the Astrophysics Explorers Program.
This work has been performed utilizing the calibration data-bases and auxiliary analysis tools developed, maintained and distributed by AstroSat-SXT and LAXPC teams with members from various institutions in India and abroad. 
C.M. is supported by an appointment with the Universities Space Research Association.
ESL acknowledges the support by DFG grant 1830Wi1860/11-1.
MTW is supported at NRL by NASA under Interagency Agreement NNG200808A.
RB acknowledges support by NASA under award number 80GSFC21M0002.

\bibliographystyle{yahapj}
\bibliography{references}
\end{document}